\documentclass{article}
\usepackage{epsfig}
\usepackage{multirow}
\usepackage{amsmath}

\def\p{\partial} 
\def\f{\frac}

\renewcommand{\thefootnote}{\fnsymbol{footnote}}
\begin{document}
\date{}

\title{Time evolution of electron flow in a model diode: Non-perturbative analysis}

\author{A.Rokhlenko and J.L.Lebowitz$^*$\thefootnote 
\medskip
\\Department of Mathematics,
Rutgers University\\ Piscataway, NJ 08854-8019}

\maketitle
\begin{abstract}

Using a combination of Eulerian and Lagrangian variables we obtain some exact results
and good approximation schemes for the time evolution of the electron flow from a 
no-current state to a final stationary current state in a planar one-dimensional diode. 
The electrons can be injected externally or generated by the cathode via field emission 
governed by a current-field law. The case of equipotential electrodes and fixed injection 
is studied along with a positive anode potential. When the current is fixed externally 
the approach to the stationary state goes without oscillations if the initial electron 
velocity is high enough and the anode can absorb the injected flow. Otherwise the 
accumulated space charge creates a potential barrier which reflects the flow and leads 
to its oscillations, but our method of analysis is invalid in such conditions. In the 
field emission case the flow goes to its stationary state through a train of decaying 
oscillations whose period is of the order of the electron transit time, in agreement 
with earlier studies based on perturbation techniques. Our approximate method does not 
permit very high cathode emissivity, though the method works when the stationary current 
density is only about $10\%$ smaller than the Child-Langmuir limit. 

\end{abstract}
\footnotetext{$^*$Also Department of Physics} 

\noindent {\bf Introduction}

\medskip
Stability of electron flows and their oscillations are of a great importance in 
technology, they attract the interest of plasma researchers and covered by numerous 
publications. While the main features of the steady flow in a diode have been known for 
a century [1], [2] an important step in studying 
time dependent states was made by Lomax [3] in 1960, who applied the 
Lagrange formulation of flow dynamics. This approach has been developed in many works, 
in particular in [4]-[8], where the authors found conditions for flow stability and modes 
of oscillations. These works all used perturbation techniques and in fact only the linear 
stability analysis describing small deviations from the steady flow have been fully 
studied. We describe here a first step toward considering far from 
stationary processes by analyzing the turn on regime in a planar diode.			

We study the electron flow in the transient state when the external parameters, such as 
injected current or acceleration voltage, are rapidly changed at $t=0$ and then stay fixed. 
We consider the space charge limited one-dimensional (1D) flow, produced by field emission
(regime I) or externally (regime II), in conditions which forbid applying linearization 
techniques [3-8]. In particular, we study the transition from no-flow initial state to a 
stationary current bearing state. The one-dimensional system should approximate a planar 
diode whose sizes in the two transversal dimensions are much larger than the inter-electrode 
distance. Our emission law is not realistic but is qualitatively of a right form, 
i.e. the stronger the cathode field the larger the emission, and it can be used 
for developing more practical models. The results cannot be quantitatively compared yet 
with the behavior of real diodes in the transient regimes and this makes it possible for 
simplicity to use dimensionless units without mapping them onto realistic ones.

\bigskip
\noindent {\bf Main equations and setup of 1D flow model}

\bigskip
The cathode is placed at $x=0$, anode at $x=x_a$, the potential, flow velocity, and 
current density are denoted as $\varphi(x,t),\ v(x,t),\ j(x,t)$ respectively as functions 
of $x$ and time $t$. We assume that the current at the cathode is determined by a function of 
the cathode electric field $f(t)$ (Regime 1) or fixed externally (Regime 2), i.e., $$
j(0,t)\equiv j(t)=F[f(t)],\ {\rm or}\ j(0,t)=j,\ \ f(t)=\f{\p \varphi}{\p x}(0,t)=E(0,t),
\ v(0,t)=\omega,\eqno(1)$$
where $E(x,t)$ is the electric field at $x$. The initial electron velocity $\omega$ is 
fixed. Let $\rho(x,t)$ is the charge density, then the current density is $j(x,t)=\rho(x,t) 
v(x,t)$, and the system is governed by the following set of equations in Euler coordinates$$
\f{\p^2\varphi}{\p x^2}(x,t)=\rho(x,t),\ \ \f{\p j}{\p x}(x,t)=-\f{\p\rho}{\p t}(x,t),\ \ 
\f{\p v}{\p t}+v\f{\p v}{\p x}=\f{1}{2}\f{\p\varphi(x,t)}{\p x}, \eqno(2)$$
setting the electron charge be $1$, its mass - $2$, and thus the electron kinetic energy is 
$v^2(x,t)$. The cathode potential $\varphi(0,t)$ is always $0$.

For $t<0$ our system is free of charge, $\rho=0$, and all its parameters have been time 
independent. The initial conditions (IC) at $t=0$ are$$
j(x,0)=0,\ \ \ \begin{cases} \varphi(x_a,0)= 0,~~\ f(x,0)=0, &{\rm Regime\ 
I},\\ 
\varphi(x_a,0)= V_a,\ f(x,0)=V_a/x_a, &{\rm Regime\ II}. \end{cases}\eqno(3)$$
\noindent
We consider processes at times $t>0$ assuming that in regime I the anode voltage at $t=0$ 
jumps to $\varphi(x_a,t)=V_a=const$, the cathode field - to $f(0,0)=V_a/x_a$, and then the 
cathode field and current will evolve as $f(0,t)\equiv f(t)$ and $j(0,t)\equiv j(t)=F[f(t)]$
respectively. In the regime II the cathode current jumps at $t=0$ from $0$ to $j(0,t)=j=const$.

Along with the Eulerian variables $x,t$ above we use the Lagrangian variables $\tau,t$ for 
an electron emitted at time $\tau$ and observed at time $t$. In this
way all flow parameters, which depend on both $x,t$, will be functions of $\tau$ and $t$.
Such functions as $f(t)$, $j(t)$, and $\varphi(0,t)$ describe the regime at the cathode 
and they depend on $t$ only. We assume that a pair $\tau,\ t$ has a one-to-one 
correspondence with some $x,\ t$. (The new external parameters for $t>0$ can be such that the
one-to-one correspondence becomes impossible and our approach fails, see later). This allows 
to find uniquely the electric field $E(\tau,t)$
as well as $\varphi(\tau,t)$ and current density $j(\tau,t)$ at the point $x,\ t$. (This 
somewhat sloppy use of the same notation in the new units will not lead to confusion). A 
straightforward analysis in these variables [3] shows that if $T$ 
is the time needed for an electron, emitted at $\tau$, to cross the diode then 
this electron is located at$$
x(\tau,t)=\int_{\tau}^t{\left[j(t')\f{(t-t')^2}{4}+f(t')\f{t-t'}{2}\right]dt'}+\omega
(t-\tau),\ \ \tau\leq t\leq T.\eqno(4a)$$
with $x(\tau,\tau +T)=x_a$. The electron velocity is the $t-$derivative of
(4a):$$v(\tau,t)=\f{\p x}{\p t}(\tau,t)= 
\omega +\f{1}{2}\int_{\tau}^t{[j(t')(t-t')+f(t')]dt'}.\eqno(4b)$$
The electric field (i.e. the force) at the point $x(\tau,t)$ is the product of the acceleration 
and mass$$
E(\tau,t)=2\f{\p^2 x}{\p t^2}(\tau,t)=f(t)+\int_\tau^t{j(t')dt'},\eqno(5)$$
where the partial derivative is taken in the Lagrange variables, i.e for fixed $\tau$.

For an electron, which hits the anode at time $t$, then Eq.(4a) takes the form$$
x_a=\int_{t-T(t)}^t{\left[j(t')\f{(t-t')^2}{4}+f(t')\f{t-t'}{2}\right]dt'}+\omega T(t),
\eqno(6)$$
where $T(t)$ is the transit time for this electron, assuming $t-T(t)\geq 0$. In the Euler  
variables the boundary condition (BC) for $\varphi(x_a,t)$ has the following form$$
V_a=\varphi(x_a,t)=\int_0^{x_a}E(x,t)dx.$$
In the Lagrangian variables using (5) it can be rewritten as$$
V_a=2\int_t^{t-T(t)}{\f{\p^2x}{\p t^2}(\tau,t)\f{\p x}{\p \tau}(\tau,t)d\tau},\eqno(7)$$
where we have used the assumption that $x$ is unique for each given $t$ and emission time 
$\tau$, therefore $dx=\f{\p x}{\p\tau} d\tau$. Eqs.(6) and (7) should hold for all $t\geq T(t)$.

One can find an explicit expression for the charge density at $\tau,t$ $$
\rho(\tau,t)=\f{\p E}{\p x}=\f{\p E}{\p\tau}\Bigg/\f{\p x}{\p\tau}=\f{j(\tau)}{j(\tau)(t-
\tau)^2/4+f(\tau)(t-\tau)/2+\omega},\eqno(8a)$$
where we have taken partial derivatives of $E$ and $x$ given in Eqs. (4),(5). We see that 
$\rho$ is always a decreasing function of $t$ if $f(t)\geq 0$ otherwise it can be 
non-monotonic. The total space charge between the electrodes at a moment $t$ in the Eulerian 
variables,$$
\varrho(1,t)=\int_0^1{\rho(x,t)dx}=E(1,t)-E(0,t).\eqno(8b)$$
This can be rewritten using Eq.(5) in the form$$
\varrho(t-T,t)=E(t-T,t)-E(t,t)=\int_{t-T(t)}^t{j(t')dt'}\eqno(8c)$$
in terms of the Lagrangian coordinates leading to simple integration of the emitted current.
We will use later an Eulerian relationship for electrons emitted at $t=0$ and located within 
the area $0<x<X(t)<1$ $$
\varrho(X,t)=\int_0^X{\rho(x,t)dx}=E(X,t)-E(0,t)=\int_{0}^t{j(t')dt'}.\eqno(8d)$$
Note that Eqs.(8b) and (8c) define the same function if $t$ is fixed. 

Evaluating the derivatives via Eq.(4) one can substitute them into (7), apply (6), and 
come to the following BC which as Eq.(6) should hold for all $t>0$ $$
V_a=x_af(t)+\int_{t-T(t)}^t{d t'\left[j(t')\f{(t-t')^2}{4}+f(t')\f{t-t'}{2}+
\omega\right]\int_{t'}^t j(t'')dt''}.\eqno(7a)$$
We see that Eqs.(6) and (7a) can be considered from now on in the usual Eulerian variables
because $t$ is the same in both systems and variable $\tau$ is absent. 

In the case when the cathode current is a function of the cathode field $j=F(f)$ (field 
emission or Child-Langmuir flow) the system (6),(8) has the following form$$
x_a=\int_{t-T(t)}^t{\left\{F[f(t')]\f{(t-t')^2}{4}+f(t')\f{t-t'}{2}+\omega\right\}d t'},$$
\vskip-0.6cm
$$\eqno(9)$$
\vskip-0.6cm
$$V_a=x_af(t)+\int_{t-T(t)}^t{d t'\left\{F[f(t')]\f{(t-t')^2}{4}+f(t')\f{t-t'}{2}+\omega
\right\}\int_{t'}^t F[f(t'')]dt''}.$$
There are two unknown functions $f(t)$ and $T(t)$ in (9) whose initial values at $t=0$ are 
determined by the IC. 

One could consider a situation with an non-zero voltage in the initial
state $V(t-)$ which makes a sudden jump $\Delta V$ at $t=0+$ and a simultaneous jump 
$\Delta V/x_a$ of the cathode field. This corresponds in our non-relativistic approximation 
to the superposition of the initial diode electric field and the homogeneous field, i.e. $E(x,0+)=E(x,0-)+\Delta V/x_a$. 

Mathematically Eqs.(9) should be sufficient for determining $f(t)$ and $T(t)$
but a clear strategy for solving them is not seen. Keeping intact the main features of a field 
emitted current we will simplify Eqs.(9) to some degree and use other functions of the flow
found above, like (4b),(8a) and (8c), to help us find an approximate solution. In particular, using
Eqs.(4b) and (8a) the time dependence of the anode current can be written as$$
j_a(t)=\f{j(t-T)v(t-T,t)}{j(t-T)T^2/4+f(t-T)T/2+\omega},\ \ T=T(t),\ t\geq T(t).\eqno(10)$$
In the stationary state, when $f$ and $j$ are constant, the denominator of (10) is the
velocity at the anode and $j_a$ is equal to the cathode current $j$ as it should. 

\bigskip\noindent
{\bf Method of solution}

\medskip
Our method of attacking this time dependent problem is to study the flow evolution at
discrete time steps which correspond to successive time intervals. If$$
\Theta_i=\sum_{k=1}^i{T_k},\ \Theta_0=0,\ i=1,2,...$$
then these intervals are $\Theta_{i-1}<t<\Theta_i$, where $i=1,2,...$. 
Here $T_i$ is the transit time of the electrons which leave the cathode at 
$t=T_{i-1}$. In particular, the first electrons starting the process at $t=0$ need $T_1$ for
their travel to the anode. When the flow can approach asymptotically to a stationary state
the most important time interval is the first one $0<t<T_1$ where the changes are dramatic,
then in our computations only several intervals $4-6$ are needed for reaching the stationary
state with a very high precision. 

This method of discrete steps can be based on Eqs.(9) or using the following scheme.
There is a possibility to derive a different set of equations in Eulerian variables implied
[10] by a relationship for $x_a=1$ $$
\f{df}{dt}(t)=\f{\p \varphi}{\p t}(1,t)+\int_0^1{[j(x,t)-j(0,t)]dx}.\eqno(11)$$ 
which fits to the situation with discrete time intervals. During the first time interval  
when the first electrons did not yet reach the anode but are located at $X(t)<1$ the current
density $j(x,t)=0$ for all $x>X(t)$. Taking into account that the anode potential is fixed 
we use Eqs.(8a) to rewrite Eq.(11) in the Lagrange variables$$
j(t)+\f{df}{dt}(t)=\int_0^{X(t)}{j(x,t)dx}=\int_t^0{v(\tau,t)\f{\p E}{\p \tau}(\tau,t)d\tau},$$
substitute $v(\tau,t)$ from (4b) and the $\tau$-derivative of $E$. The result is $$
j(t)+\f{df}{dt}(t)=\omega\int_0^t{j(t')dt'}+\f{1}{2}\int_0^t{dt'[j(t')(t-t')+f(t')]
\int_0^{t'} {j(t'')dt''}}.\eqno(12)$$
This provides a closed equation for finding $f(t)$ when $j(t)$ is given or $j(t)=F(f)$ and 
$t\leq T_1$. Equations corresponding to (12) for the time intervals $\Theta_{i-1}<t<\Theta_i$ 
can be useful only for $j(t)=const$ for better precision when $t>T_1$. We apply here a more 
universal scheme with Eqs.(9) which provide sufficiently accurate results.

This approximation is based on representing the function $f^i(t)$ on each time interval as
a series, 
$$f^i(t)=\sum_{n=0}{c_n^i\left(\f{t-\Theta_{i-1}}{T_{i}}\right)^i},\eqno(13)$$
and truncating it to a finite sum with $n\leq N$. The number of terms $N$ will be chosen in 
accord with the available number of equations for determining $c_n^i,\ n=0..N$ and the 
corresponding transition time $T_i$. 

\bigskip\noindent
{\bf Regime I with linear $F[f]$, $\omega=0$, and $V_a(t>0)=1$}

\medskip
Let $F[f(t)]=af(t),\ a>0$ and at $t=0-$ we have $x_a=1$ and in the Eulerian variables
$\varphi(x,0)=0,\ \rho(x,0)=0$ are changed with$$
\varphi(x,0)=x,\ \ f(x,0)=1\ {\rm for}\ t=0+,\eqno(14)$$
In this way we study the flow after the diode is turned 
on by applying the voltage $V_a=1$ suddenly. When $\omega\neq 0$ all our equations become only 
slightly more complicated, we consider the simplification $\omega =0$ keeping in mind mainly 
high applied voltages when $\omega$ can be neglected.

Thus we come to a problem of finding two unknown functions $f(t)$ and $T(t)$ which satisfy
Eqs.(9) in the form$$
\f{1}{4}\int_{t-T}^{t}{f(t')[a(t-t')^2+2(t-t')]d t'}=1,\eqno(15a)$$
$$f(t)+\f{a}{4}\int_{t-T}^{t}{dt'f(t')[a(t-t')^2+2(t-t')]
\int_{t'}^{t}f(t'')dt''}=1\eqno(15b)$$
for all $t\geq T$ where $T$ is the transit time $T=T(t)$ for the electron which hits the
anode at time $t$. Here as before $f(t)$ is the electric field at the emitter. Note that
Eq.(15a) is already implemented in Eq.(15b), as well as in (17b) below.

Eqs.(9) or (15) with proper BC and IC obviously uniquely determine the flow evolution in
our system because they are valid for all $t\geq T(t)$ in (15), but finding $f(t)$ from them 
is a difficult problem. Approximate methods of solving them in our approach 
will involve the use of discrete sets of values of $t$, which 
leads to losing some information. As Eqs.(15) are identities valid for all $t$ we 
supplement Eqs.(15) with three derivative of these equations. Thus we conserve at least some 
information to get additional identities of the same nature as (15). The derivative of the 
integral in (15a) is zero and therefore$$
\left(1-\f{dT}{dt}\right)f(t-T)(aT^2+2T)=2\int_{t-T}^t{f(t')[a(t-t')+1]dt'}.\eqno(16)$$
This equation will be used for evaluating the derivative $T'(t)$ in computations below.
After some manipulations one can derive from (15b) and (16) the following relation$$
\f{df}{dt}+af(t)=\f{a}{2}\int_{t-T}^t{dt'f(t')[a(t-t')+1]\int_{t-T}^{t'}{f(t'')dt''}},
\eqno(17a)$$
which as well as (15) is valid for all $t\geq T(t)$ but does not involve the derivative  
$T'(t)$ explicitly. We add to Eqs.(15) and (17a) the derivative of Eq.(17a) 
$$\f{d^2f}{dt^2}(t)+a\f{df}{dt}(t)=\f{af(t)}{2}\int_{t-T}^t{f(t')dt'}+\f{a^2}{4}\left[
\int_{t-T}^t{f(t')dt'}\right]^2$$
$$-\f{a}{aT^2+2T}\left\{\int_{t-T}^t{f(t')[a(t-t')+1]dt'}\right\}^2,\eqno(17b)$$
which can be rewritten using Eqs.(4b) and (8c) as$$
\f{d^2f}{dt^2}(t)+a\f{df}{dt}(t)=\f{f(t)}{2}\varrho(t)+\f{\varrho^2(t)}{4}-
\f{4a}{aT^2+2T}v^2(t).$$
Then we calculate the time derivative of Eq.(17b) and obtain the following relation in 
the Eulerian variables 
$$\f{d^3f}{dt^3}(t)+a\f{d^2f}{dt^2}(t)-\f{\varrho(t)}{2}\f{df}{dt}(t)=[f(t)+\varrho(t)]\left[
\f{af(t)}{2}-\f{6av(t)}{aT^2+2T}\right]$$
$$+8av^2(t)\f{aT+1}{(aT^2+2T)^2}\left[3-\f{4v(t)}{aT^2+2T}\right]\eqno(17c)$$
for the moment $t=T(t)$ when the electrons emitted at $t=0$ reach the anode.

The set of five Eqs.(15) and (17) will be used in our calculations. Using Eq.(4b) the equation 
(10) for the anode current density can be now simplified to the form$$
j_a(T)=\f{4av(T)}{aT^2+2T},\eqno(18)$$
where the electron velocity at the anode $v$ and $T$ are both functions of a single variable 
$t$ and can be treated in the Eulerian units. For this one should calculate $T(t)$ by the 
Lagrangian approach to find $j_a$ as a continuous function of time. Eqs.(16) and (18) imply 
a useful relationship $j_a(T)=af(0)(1-T')$ for the anode current.

In the stationary case $f(t)=f=const$ the electron velocity at the anode $v_a=1$ and 
Eqs.(8a), (17a) imply $j=af=v_a\rho_a=\rho(T+\tau,\tau)=4a/(aT^2+2T)$. Thus $f=4/(aT^2+2T)$
and from Eq.(6) $f=12/(aT^3+3T^2)$. Therefore$$ 
T_{st}=\f{4}{\sqrt{1+2a/3+a^2}-a+1},\ \ f_{st}=\f{1+3a^2}{2}+\f{1-3a}{2}\sqrt{1+2a/3+a^2},$$
and, in particular, the well known results $T=3,\ j=4/9,\ f=0$ if $a=\infty$ for the CL flow, 
and $T=2$ when $a=0$ (no current). These formulas are in agreement with corresponding relations 
found in [9] and [8].

On the initial time interval $0<t<T_1$ using the approximation (13) with a finite $N$ we have$$
f(T_1)=\sum_{i=0}^N{c_i^1},\ \varrho(T_1)=aT\sum_{i=0}^N{\f{c_i^1}{i+1}},\ \ 
v(T_1)=\f{T_1}{2}\sum_{i=0}^N{\f{c_i^1}{i+1}\left(1+\f{aT_1}{i+2}\right)}.$$
Using these notations the straightforward but tedious calculations allow to rewrite Eqs.(15a), 
(15b), and (17a) - (17c) respectively in the following form:$$
\f{1}{2}\sum_{i=0}^N{\f{c_i^1T_1^2}{(i+1)(i+2)}\left(1+\f{aT_1}{i+3}\right)}=1,\eqno(19a)$$
$$f(T_1)+\varrho(T_1)
-\f{a}{2}\sum_{i,j=0}^N{\f{c_i^1c_j^1T_1^3}{(i+1)(i+j+2)(i+j+3)}\left(1+\f{aT_1}{i+j+4}\right)}=1, 
\eqno(19b)$$
$$2\sum_{i=0}^N{c_i^1\left(1+\f{i}{aT_1}\right)}=\sum_{i,j=0}^N{\f{c_i^1c_j^1T_1^2}{(i+1)(i+j+2)}\left(
1+\f{aT_1}{i+j+3}\right)},\eqno(19c)$$
$$\f{1}{T_1}\sum_{i=0}^N{c_i^1i\left(1+\f{i-1}{aT_1}\right)}
=\f{2f(T_1)+\varrho(T_1)}{4a}\varrho(T_1)-\f{4v^2(T)}{aT_1^2+2T}.\eqno(19d)$$

$$8\f{v^2(T_1)(aT_1+1)}{(aT_1^2+2T_1)^2}\left[3-\f{4v(T_1)}{aT_1^2+2T_1}\right]+[f(T_1)+\varrho(T_1)]
\left[\f{f(T_1)}{2}-\f{6v(T_1)}{aT_1^2+2T_1}\right]$$
$$=\f{1}{aT_1^3}\sum_{i=0}^N{c_i^1 i(i-1)(i-2+aT_1)}-\f{\varrho(T_1)}{2aT}\sum_{i=0}^N{ic_i^1}.\eqno(19e)$$
Eqs.(19) are five algebraic equations for $f(t)$ and $T_j$ on each consequent interval 
$\Theta_{j-1}\leq t\leq T_j+\Theta_{j-1}$, and by differentiation one can have as many equation as he wants if their practical use is reasonable. Our method of using a polynomial form (13) neglects the
higher derivatives of $f(t)$ and we think that to go beyond $d^3f/dt^3$ (in Eq.(17c)) is not
sensible.

Note that Eq.(12) can be rewritten using the expansion (13) of $f(t)$ on the first interval 
$0\leq t<T_1$ as following$$
\sum_{i=0}{\f{c_i^1}{T_1^i}(at^i+it^{i-1})}=a\sum_{i=0}{\f{c_i^1}{T_1^i}\f{t^{i+1}}{i+1}\left[
\omega +\f{1}{2}\sum_{j=0}\f{c_j^1t^{j+1}}{T_1^j(i+j+2)}\left(1+\f{at}{i+j+3}\right)\right]}.$$
This relationship and the BC make it possible to evaluate all the coefficients of the 
series (13). The first of them are $$
c_0^1=1,\ \f{c_1^1}{T_1}=-a,\ \f{c_2^1}{T_1^2}=\f{\omega a+a^2}{2},\ \f{c_3^1}{T_1^3}=a\f{1-4\omega a-2a^2}{12},...\eqno(20)$$
and the computation of the higher ones is more laborious but straightforward. The general term 
$b_k=c_k^1/T_1^k$ can be computed from the following equation$$
(n+1)b_{n+1}=-ab_n+a\omega\f{b_{n-1}}{n}+\f{a}{4}\sum_{i=0}^{n-2}\f{b_ib_{n-2-i}}{(i+1)(n-1-i)}
+\f{a^2}{4n}\sum_{i=0}^{n-3}\f{b_ib_{n-3-i}}{(i+1)(n-2-i)}.$$
This provides an important information about the first coefficients $c_0^1,c_1^1,...$, but
the convergence of the series is very slow in this case (Regime I). We use only these
two coefficient then model $f(t)$ with a finite sum and apply Eqs.(19) for this model of the
field emission.

Two more remarks. A simple analysis of Eqs.(1),(2) allows also to find the upper bound for 
$T_1$ in the general case. By integrating twice the Maxwell equation in (2) and using BC 
we have first$$
\f{\p \varphi}{\p x}(x,t)=f(t)+\int_0^x{\rho(x',t)dx'}\eqno(21a)$$
and then the potential in the form$$
\varphi(x,t)=xf(t)+\int_0^x{(x-x')\rho(x')dx'}.\eqno(21b)$$
Let us consider the moving boundary $X(t)$ between the space charge and vacuum and thus 
$\rho(x>X)=0$. Now we exhibit Eq.(21b) at the anode $x=1$ and Eq.(21a) at $x=X(t)$ $$
1=f(t)+\int_0^{X(t)}{(1-x')\rho(x')dx'},\ \ \f{\p \varphi}{\p x}(X,t)= 
f(t)+\int_0^{X(t)}{\rho(x',t)dx'}.\eqno(21c)$$
Combining two Eqs.(21c) we derive$$
\f{\p \varphi}{\p x}(X,t)=1+\int_0^{X(t)}{x'\rho(x',t)dx'}.\eqno(22)$$
Eq.(22) shows explicitly that the frontal electrons are always accelerated by the field 
stronger than the vacuum field, i.e. $1$, and therefore the crossing time $T_1$ is less
than $2$, which is consistent with the intuition. One more important statement relates to
the continuity of the cathode field at $t=T_1$. As $\varphi(1,t)$ is constant while the 
cathode field $f(t)$ and current $j(0,t)$ are continuous functions the derivative $df/dt$ 
in Eqs.(17) are continuous at $T_1$ because $X(t)$ approaches and stays equal 
to $1$. This justifies our technique of solving Eqs.(17) on the second and the following 
step and keeping continuity of $f(t)$ with its first derivative at $t=T_j$. It is easy 
to see that the second derivative of $f$ at $T_1$ is discontinuous in general.

\bigskip

\noindent {\bf Flow evolution for different cathode emissivity}
\bigskip

Using the five Eqs.(19) for $0<t<T_1$ and Eqs.(20) we evaluate $c_3^1,...,c_6^1$ and $T_1$.
Then we will study the flow on the interval $T_1<t\leq T_1+T_2$, where $T_2$ is the 
transition time for electrons emitted at $t=T_1$, and so on step by step until the current
stabilization. On each subsequent interval the IC for $f(T_i)$ and its first derivative will 
be matched with the previous values found by calculations. Our 5 equations are Eqs.(17a)...(17e), 
this implies that on all intervals we can evaluate corresponding $T_i$ and four coefficients 
$c_2^i...c_5^i$ because two coefficients $c_0^i$ and $c_1^i$ are determined by the IC at 
$t=\Theta_{i-1}$ mentioned above. The technique of using Eq.(19) can not be applied on the 
later intervals because the upper limit of the integral in (22) must be equal to $1$ there. 
It turns out that relatively small current variations at later time $t>T_1$ are well described 
without using higher derivatives of $f(t)$.

Taking $a=3$ this scheme can be realized and the time dependence of the cathode current 
density is exhibited in Fig.1. 

\vskip0.3cm\hskip0.2cm
\epsfig{file=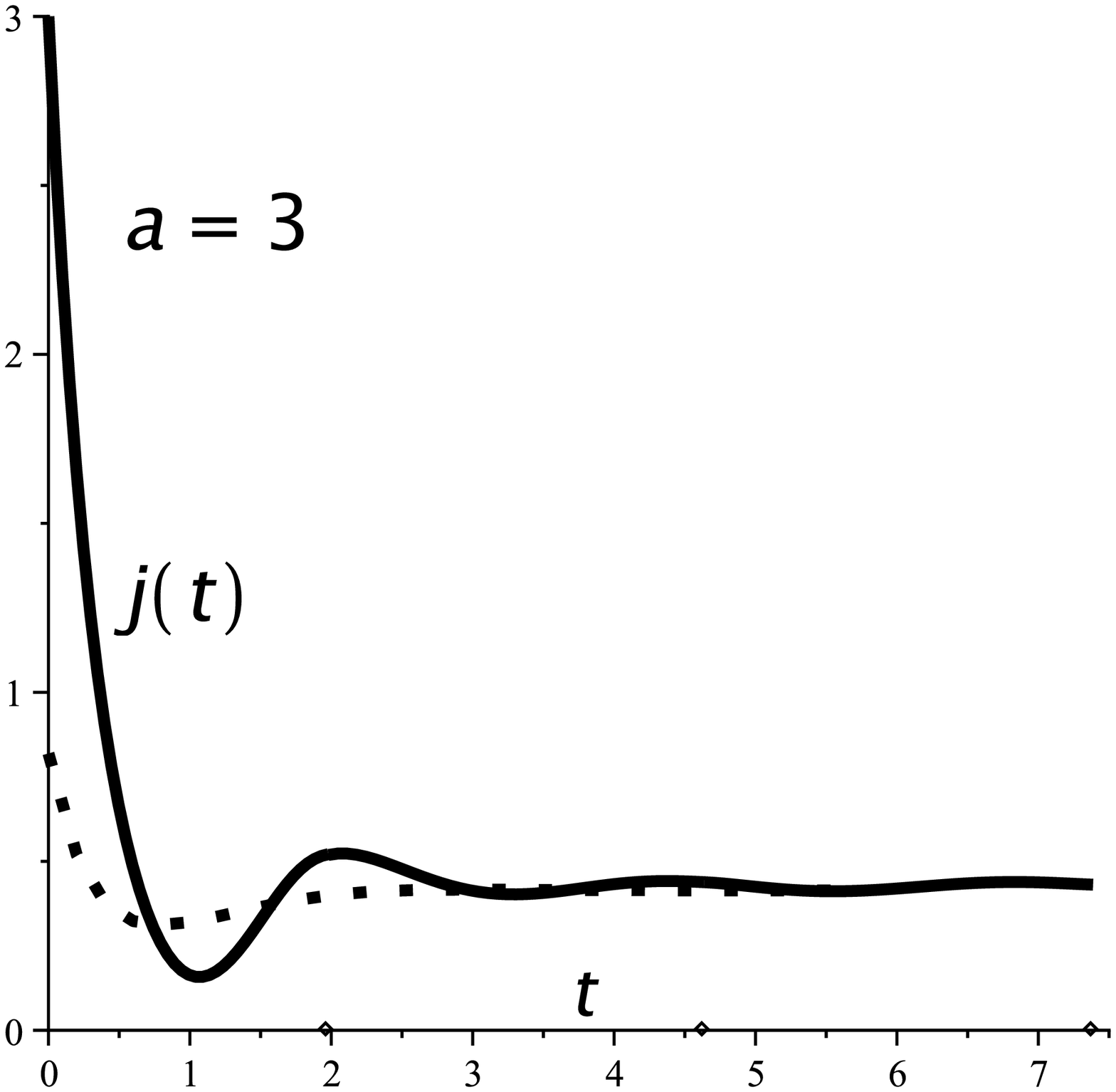, width=10cm, height=7cm} 

\vskip-6.5cm\hskip4cm
\epsfig{file=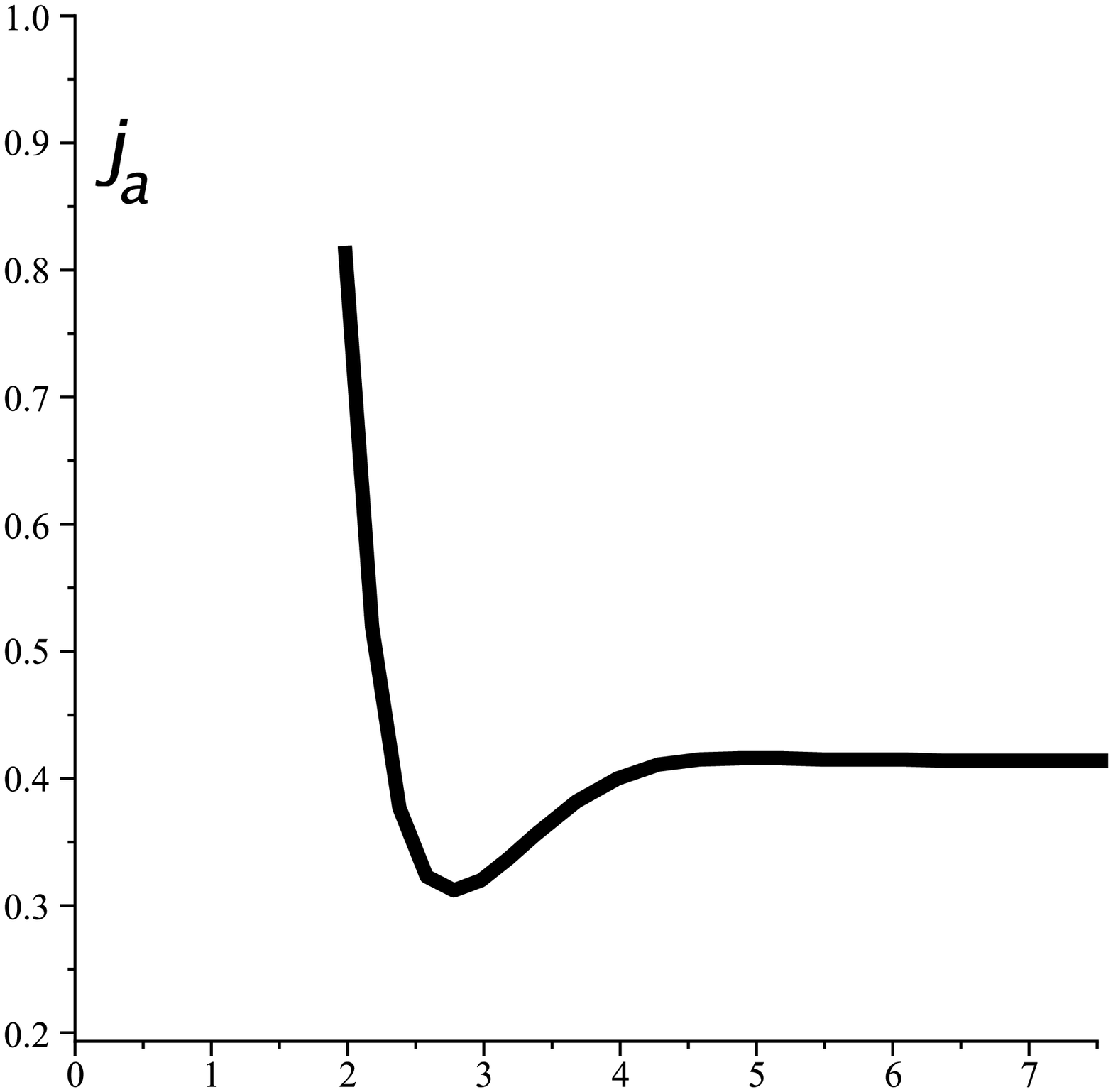, width=5cm, height=4cm} 

\vskip2.3cm
\centerline{\small FIG.1. Evolution of cathode and anode currents for $a=3$. Inset shows anode current $j_a$}

{\small Dotted curve for $j_a$ shifted left by $T_1$. On $t$-axis are shown points for $T_1,
\ \Theta_2,\ \Theta_3$} 

\medskip
The stabilization process almost finishes after $t>T_1+T_2$
and $j(t)$ asymptotically approaches to the stationary [9] value. For $0\leq a\leq 3.5$ 
the asymptotic values of $j$ for $a=1,\ 2, \ 3,\ 3.5$ are found at corresponding 
$t=T_1+T_2+T_3$ to be $0.367,\ 0.417,\ 0.432,\ 0.434$ respectively. They differ from the 
currents in the steady states at most by $0.4\%$ for $a=3.5$. But unfortunately larger values 
of $a$ cannot be considered by this approximation because the minimum current, see Fig.1, 
becomes negative. This contradicts the initial assumption and physics: the flow becomes $0$ 
when $f\leq 0$, i.e. ($j\neq af$). The minimum of $j(t)$ at $t\approx 0.1$ in Fig.1, $0.02$, 
is already close to zero. Note that the time of the flow stabilization is in agreement 
with the linear theory [7] for small perturbations of the stationary state. Using Eq.(10)
we exhibit also the anode current evolution which clearly starts at $t=T_1$.
On the first step the transition time $T_1=1.98$ is close to $2$, which corresponds to the case  
when an electron moves without space charge in the same diode. Surprisingly $T_1$ is quite close 
to $2$ though the space charge pushes forward the first bunch of electrons while the anode 
attracts them. Starting from the second step all $T_j$ stay close ($T_2=2.66,\ T_3=2.75$)
to the asymptotic stationary value $T_{as}=2.732$ [9] as well as the current density $j_{as}\to 
0.431$. These quantities for $a=3$ are not far away from the Child-Langmuir ones: $T_{CL}=3,
\ j_{CL}=0.444$ when the emissivity is infinite, $a=\infty$. 

Using Eq.(10) one can find the time dependence of the anode current which is shown in the 
inset of Fig.1 for the case when $a=3$. It appears only for $t\geq T_1$. We plot the cathode 
current together with the anode current shifted for the sake of illustration to the left by 
$T_1$ (the electrons emitted at $t=0$ reach the anode at $t=T_1$). In order to have a sufficient
number of points for $j_a(t)$ we solve Eqs.(15) not only for $\tau=0$ and $\tau=T_j,\ j=1,2,...$ 
but also in intermediate points with smaller intervals like $\tau_k =0.1k,\ k=1,2,...$. This 
allows to find also corresponding values for the cathode current $j$ and $T$. 

The asymptotic behavior of both graphs in Fig.1 is identical. Interestingly the initial 
anode current is lower than the cathode one as a dense space charge is formed temporarily in 
front of the electrons which were emitted later. It decreases their acceleration and thus makes 
slower their speed near the anode.

\hskip2cm
\epsfig{file=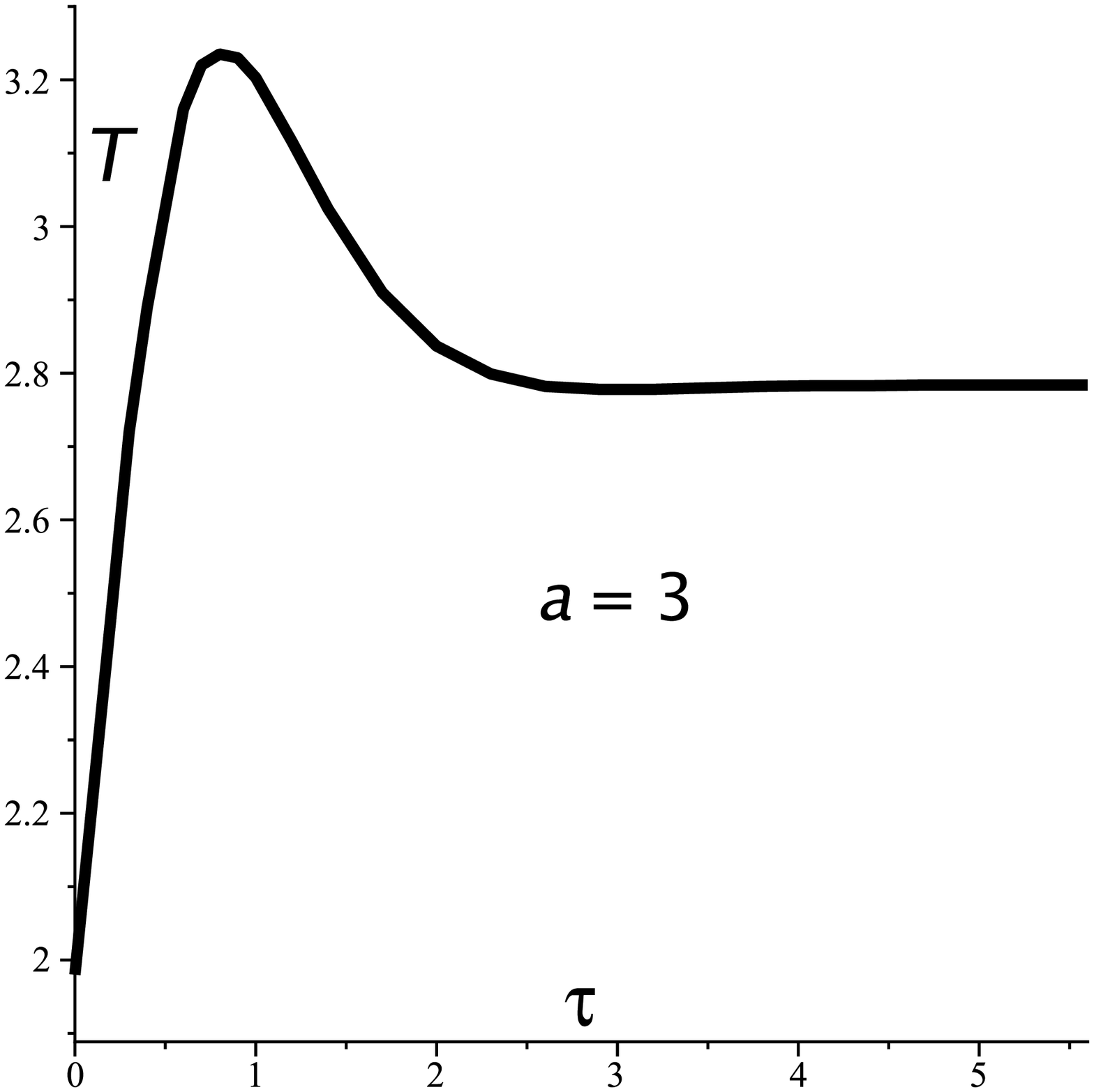, width=7cm, height=6cm} 

\vskip0.2cm
\centerline{\small FIG.2. Transition time $T$ for $a=3$ as function of $\tau$ for $0<\tau<5.6$}

\medskip
In Fig.2 is shown the time dependence of the transition time $T$ starting from $T=T_1$. The 
frontal electrons move a little faster than in vacuum and they push back the layer of the space 
charge following behind them, thus it moves even slower than in the stationary conditions. 

\vskip0.3cm\hskip0.2cm
\epsfig{file=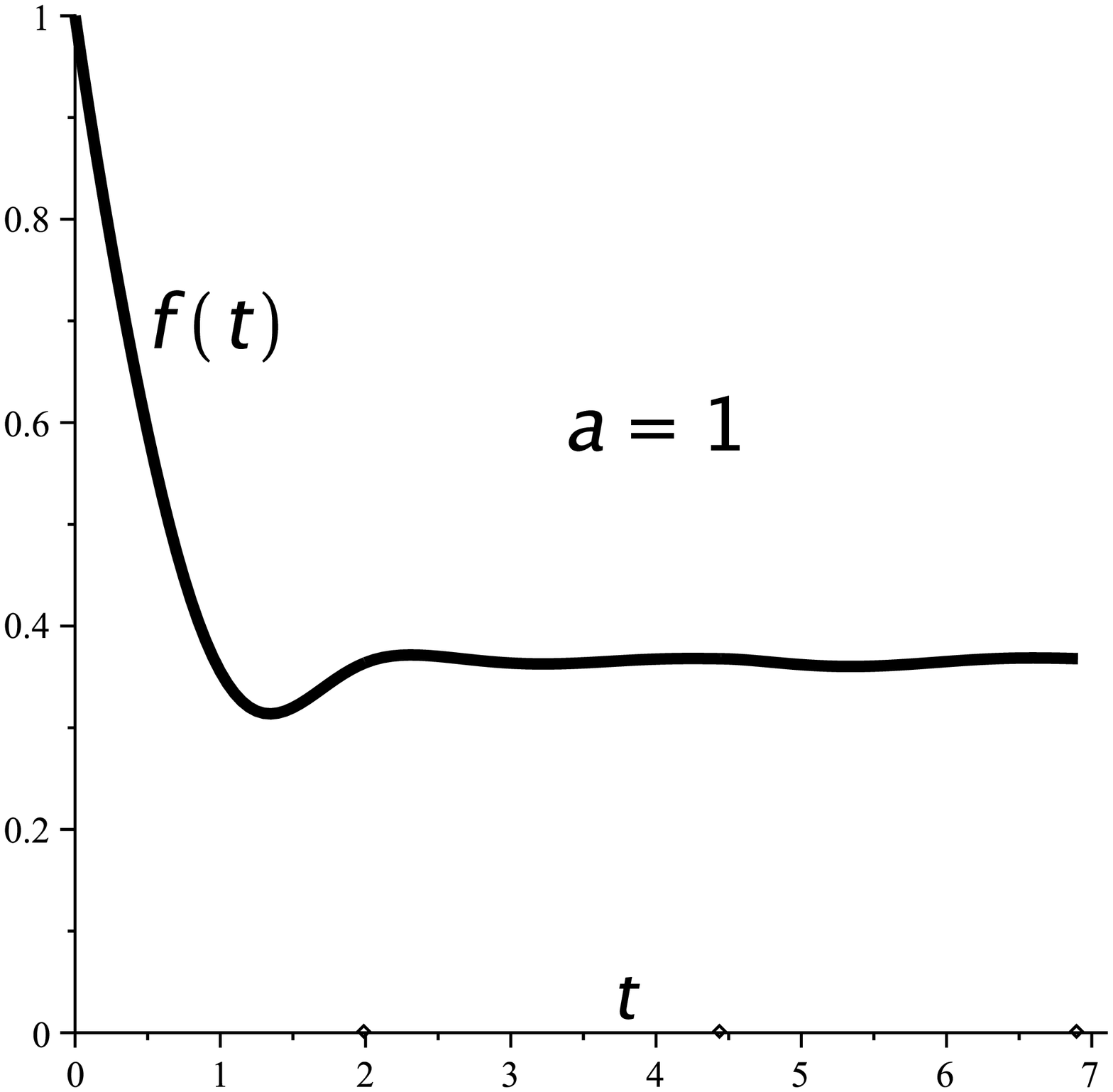, width=8cm, height=6cm} 

\vskip0.2cm
\centerline{\small FIG.3. Evolution of cathode field for $a=1$ on interval $0<t<\Theta_3$}
\centerline
{\small Points on $t$-axis correspond to $T_1=1.998,\ T_2=2.449,\ T_3=2.458.$} 

\medskip\noindent
Fig.3 exhibits the flow properties when $a=1$. The stationary values in this case are $T_a=2.449$ 
and  $j_a\to 0.367$ while the oscillations are rather weak, i.e when $a<1$ the diode comes to its 
stationary state with a short delay (about one half of the transition time in free space) and 
almost monotonically. 

\bigskip

\noindent {\bf Regime II with fixed injected current}

\bigskip\noindent	
This setup is much simpler than the case of the field emission (even for the present emission
model) because Eq.(12) becomes linear for the unknown function $f(t)$. In this case Eqs.(9) 
and (19) are valid, they are linear for $c_n^i$ and therefore much simpler.

\medskip
1. $V_a=1$

Let us substitute $j=Const$ into Eq.(12) and perform the integration$$
\f{df}{dt}(t)-\f{j}{2}\int_0^t{t'f(t')dt'}=-j+\omega jt+\f{j^2t^3}{12}.\eqno(23)$$
Eq.(23) is exact for $t\leq T_1$. It can be solved (see Supplement) explicitly with the help 
of Bessel functions, but the solution in the form of a series similar to Eq.(16)$$
f(t)=\sum_{n=0}{b_nt^n}$$
provides excellent precision and a faster numerical procedure. We substitute it into Eq.(23), 
apply the IC $f(0)=1$ and obtain the following relations which
define the cathode field on the first time interval $$
b_0=1,\ b_1=-j,\ b_2=\f{\omega j}{2},\ b_3=\f{b_0j}{12},\ b_4=-\f{j^2}{48},\ b_{n+3}=\f{jb_n}{2(n+2)(n+3)},\ (n\geq 2).$$
The series converges rapidly for not very large $j$ and $\omega$, and 12-15 terms guarantee 
a good precision but we used about 30 terms. 

For $t>T_1$ one needs to construct the following set of equations for $t=T(t)$ analogous to 
Eqs.(17) for each consequent step$$
\int_{t-T}^t{\left[j\f{(t-t')^2}{4}+f(t')\f{t-t'}{2}+\omega\right]dt'}=1,\eqno(24a)$$
$$f(t)+j\int_{t-T}^t{\left[j\f{(t-t')^3}{4}+f(t')\f{(t-t')^2}{2}+\omega(t-t')\right]dt'}=1,
\eqno(24b)$$
$$\f{df}{dt}(t)-\f{jT}{2}\int_{t-T}^t{f(t')dt'}+j\int_{t-T}^t{f(t')(t-t')dt'}=0,\eqno(24c)$$
$$\f{d^2f}{dt^2}(t)+j\f{[Tf(t-T)-\int_{t-T}^t{f(t')dt'}]^2}{jT^2+2Tf(t-T)+4\omega}= 
jT\f{f(t)+f(t-T)}{2}.\eqno(24d)$$
Using the approximation (16) Eqs. (24) are converted into a set similar to Eqs.(19)$$
\f{jT^3}{12}+\f{T^2}{2}\sum_{i=0}^N{\f{c_i}{(i+1)(i+2)}}+\omega T=1,\eqno(25a)$$
$$\sum_{i=0}^N{c_i}+\f{j\omega T^2}{2}+\f{j^2T^4}{16}+jT^3\sum_{i=0}^N\f{c_i}{(i+1)(i+2)(i+3)}
=1,\eqno(25b)$$
$$\sum_{i=1}^N{ic_i\left[1-\f{jT^3}{2(i+1)(i+2)}\right ]}=0,\eqno(25c)$$
$$\sum_{i=2}^N{i(i-1)c_i}=\f{jT^3}{2}\left(f_0+\sum_{i=0}^N{c_i\f{i-1}{i+1}}\right)-\f{jT^4} 
{4\omega +2Tf_0+jT^2}\left(f_0-\sum_{i=0}^N{\f{c_i}{i+1}}\right)^2,\eqno(25d)$$
where $f_0$ denotes the cathode field at the start of the corresponding step.

The results of such computation for $\omega=0.5$ with $j=1$ and $j=1.5$ are shown in Fig.4. 
The current becomes stable without oscillations in this case and after 5-8 transition times 
(which vary approximately from $1.2$ to $1.5$) the cathode field differences from the 
quantities, evaluated in [7] for the stationary flows, are about $10^{-4}\%$.

\vskip0.3cm\hskip0.2cm
\epsfig{file=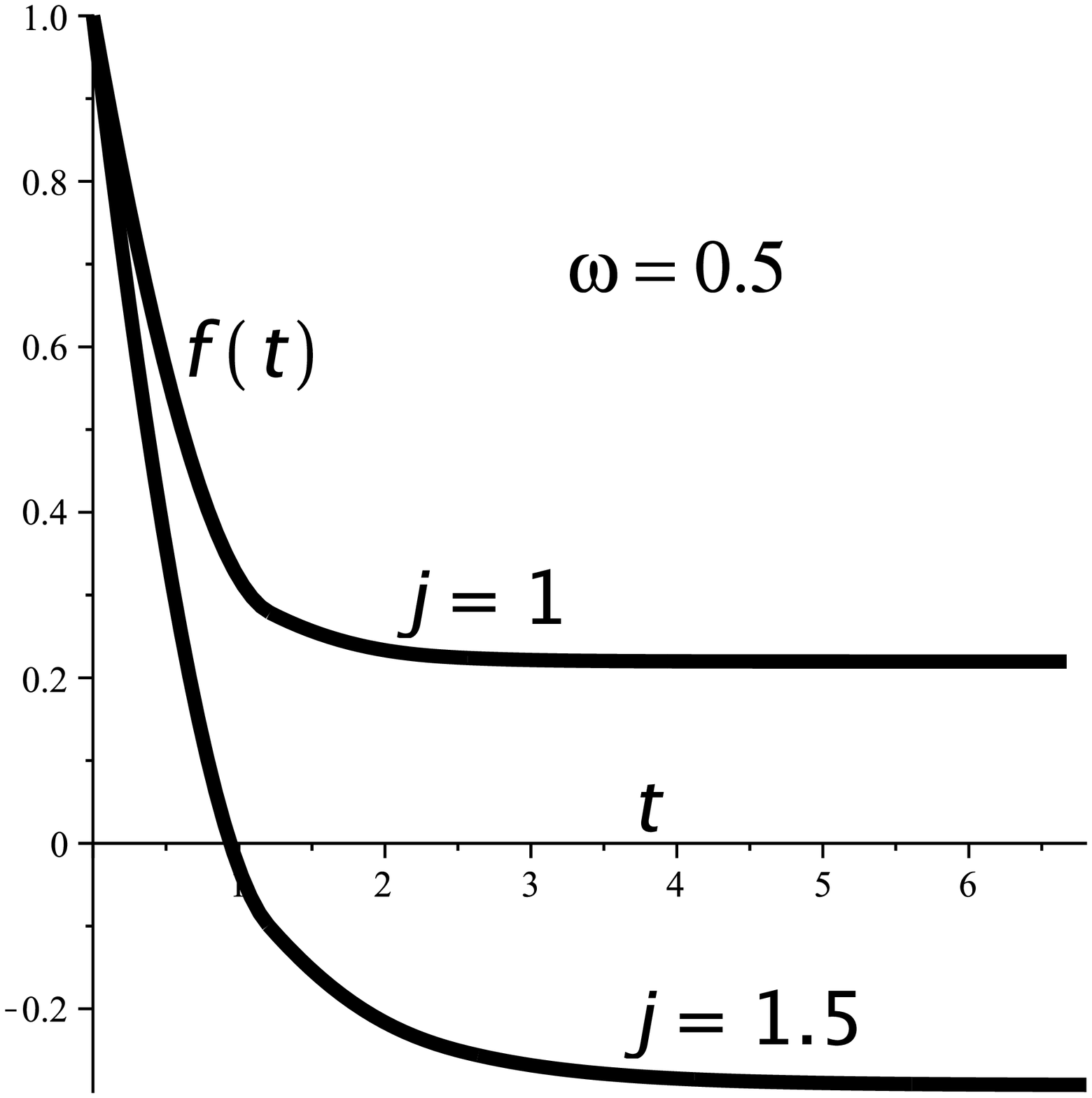, width=8cm, height=6cm} 

\vskip0.2cm
\centerline{\small FIG.4. Cathode electric field evolution when $j=1$ and $j=1.5$ with $\omega=0.5$}
\medskip\noindent
When $j=1$ there is no virtual cathode and $f>0$, but with the same $\omega$ the stronger 
current $j=1.5$ creates a virtual cathode and $f$ becomes negative. The transition times on the 
consequent steps are $1.211,\ 1.355,\ 1.367,\ 1.368,\ 1.368$ for $j=1$ and $1.201,\ 1.437,\ 1.482,\ 
1.491,\ 1.493$ for $j=1.5$ and they practically become stabilized later. The initial anode current 
densities at $t=T_1$ are $0.814$ and $1.127$ for $j=1$ and $1.5$ respectively. Note that the
frontal electron velocities at the anode are $1.199$ and $1.234$, i.e. larger than they would
be in vacuum $\sqrt{1+\omega^2}=1.118$, while the anode current densities are lower than 
corresponding values of $j$.  The frontal electrons in vacuum would need $T=1/\omega$ for 
crossing the diode, but on the first interval $t<T_1$ they move faster being pushed by the 
field of particles behind them.  

\bigskip\noindent
2. $V_a=0$

\medskip	
We consider now the closed system with equipotential electrodes. The injected current creates
the space charge and the lowest (negative) potential - virtual cathode - in the diode occurs 
somewhere at $0<x<1$.   

\vskip0.3cm\hskip0.2cm
\epsfig{file=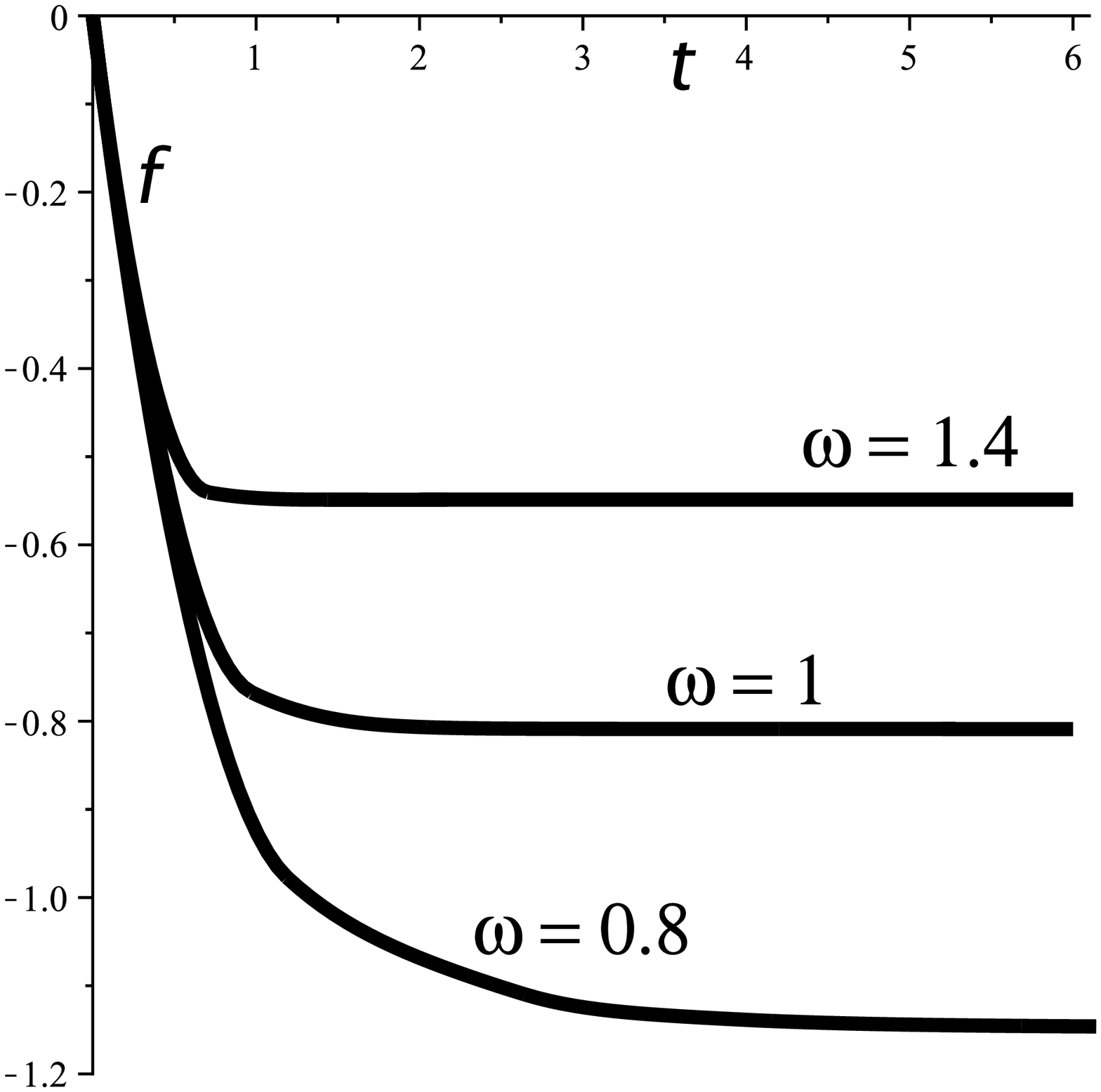, width=8cm, height=6cm} 

\vskip0.2cm
\centerline{\small FIG.5. Cathode electric field evolution when $j=1.5$ and $\omega=0.8,\ 1,\ 1.4$}

\medskip\noindent
In Fig.5 we show the cathode field behavior when the injected current is fixed for the particle 
initial velocity $\omega$ of three different values $0.8,\ 1.0$ and $1.4$. It is clear that the 
larger is $\omega$ the smaller is the charge density and when $\omega=0.7$ we did not reach the flow 
stabilization because the anode cannot absorb the incoming flow and probably the charge density 
inside the diode grows infinitely, there are possible also some chaotic behavior which is beyond 
our present computational techniques. Clearly the same can happen also with a positive anode when
the current density is large while $\omega$ is small. 

For $\omega=0.8$ the field $f$ tends to $-1.15$, it becomes
closer to zero when $\omega$ increases. The cathode electric field always goes to its asymptotic 
value without oscillations, see Fig.5, and gets stabilized relatively fast. The values of 
transition times $T_k, k=1..n$ vary from $0.8$ to $1.3$, there are about five of them in Fig.5. 
Electrons after entering the diode meet an already formed space charge and transition times get 
longer depending on its density.

In Fig.6 is shown the behavior of the space charge boundaries $X(t)$ and the locations of the
virtual cathode $y(t)$ for $\omega=0.8$ and $1.4$ on the corresponding first time intervals $0<t<T_1$ ($\approx 1.19$ and $0.72$ respectively).

\vskip0.3cm\hskip-0.5cm
\epsfig{file=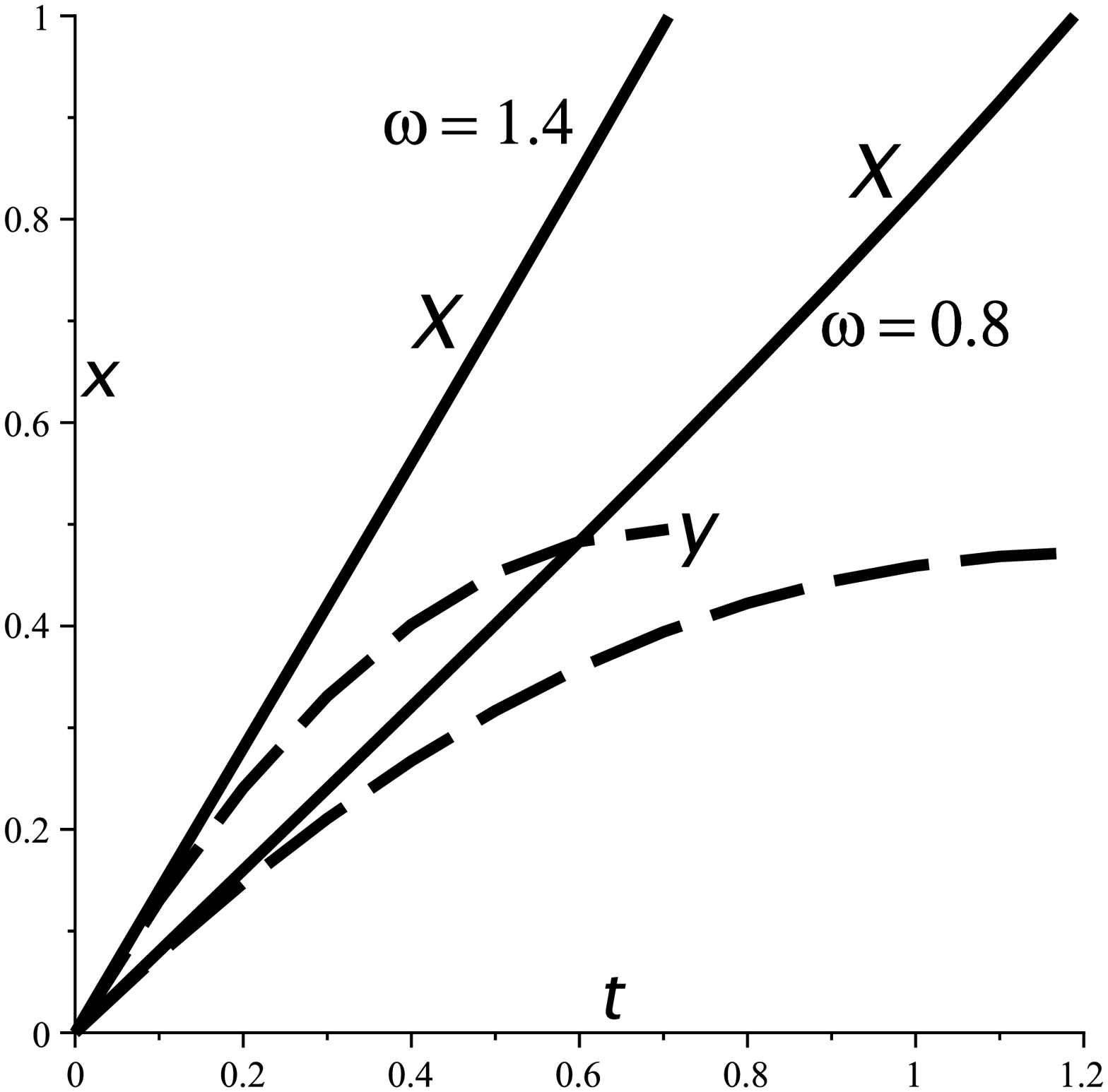, width=6cm, height=6cm} 

\vskip-6cm\hskip5.5cm
\epsfig{file=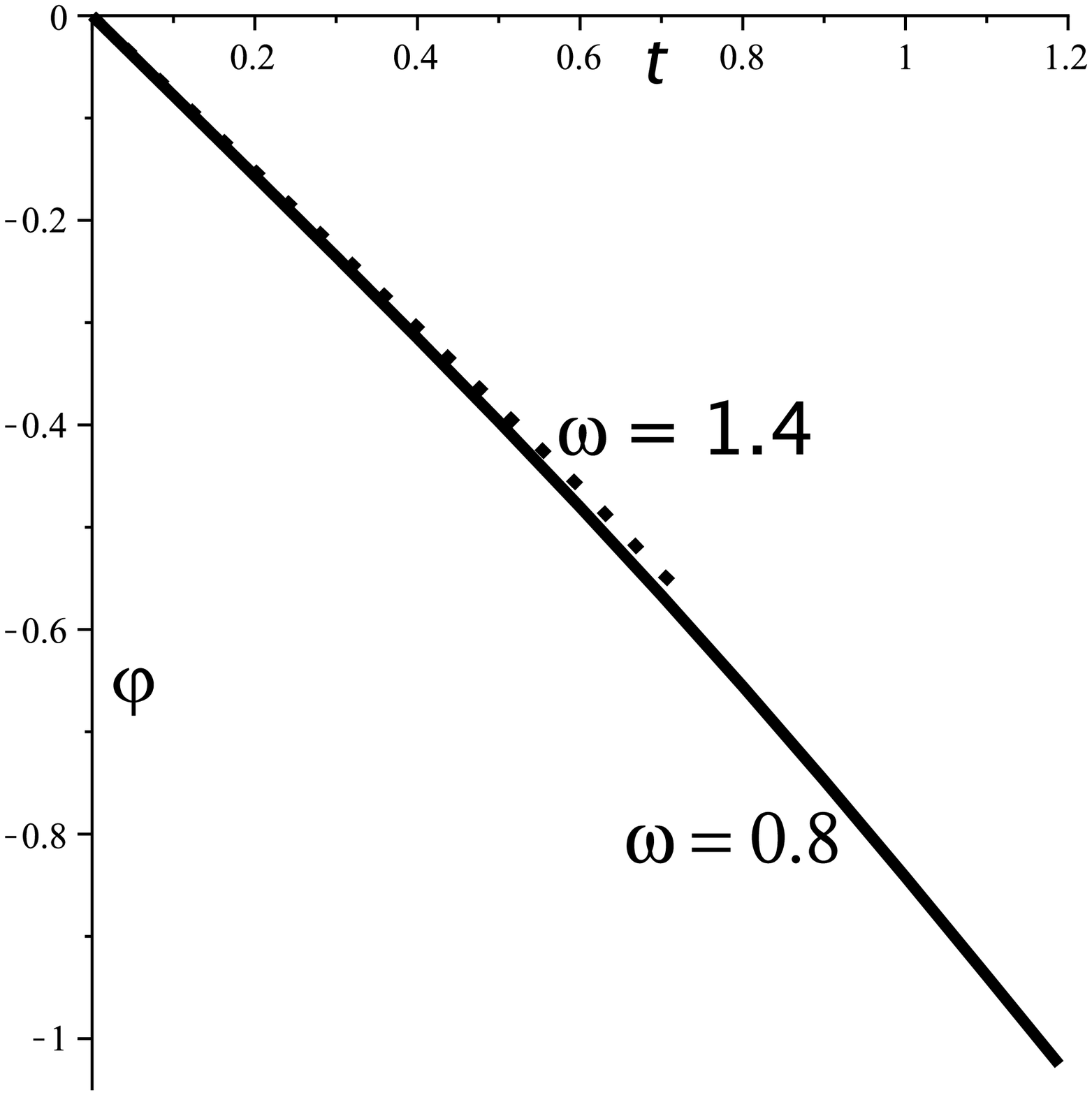, width=6cm, height=6cm} 
\vskip0.2cm
\hskip-0.2cm{\small FIG.6. Plots of $X(t)$, $y(t)$ (left) and $\varphi(y,t)$ (right) when $j=1.5$ and $\omega=0.8,\ 1.4$}

\medskip\noindent
As seen in Fig.6 location $X(t)$ of the frontal electrons moves almost by a linear in $t$ law 
$X(t)\approx \omega t$, the maximum deviation is only a few percents. The construction of Fig.6 
is simple:
the minimum corresponds to zero electric field and Eq.(5) allows to evaluate $\tau$ for each
$t\leq T_1$ as soon as $f(t)$ is computed. Then $y$ is evaluated by Eq.(4a) which also is used
for $X(t)$ when $\tau$ is replaced by zero. A similar graph can be made for the case $V_a=1$
above when the virtual cathode is realized there and $f(t)<0$, say for the case illustrated by
the lower curve in Fig.4.

The evolution of the potential of the virtual cathode is shown in right figure.
For smaller initial velocity the space charge is denser and thus the potential is more
negative. On the next time intervals the potential goes deeper and approaches to 
its stationary value. In Fig.6 we see that a larger initial velocity $\omega$ makes $T_1$ 
shorter and curves stop at corresponding values $T_1$. 

\bigskip
\noindent {\bf On the validity of the method used}
\bigskip				

We present now a simple sufficient condition which roughly indicates the limitation for the 
method used here of analyzing the flow when $j$ is fixed externally. 
Let us consider at time $t$ two electrons emitted at $\tau_1$ and 
$\tau_2>\tau_1$ and assume that $x(\tau_1,t)\leq x(\tau_2,t)$, i.e. a later emitted electron
can overcome an earlier emitted one. Thus $Eq.(4a)$ implies for them$$
x(\tau_1,t)-x(\tau_2,t)=\int_{\tau_1}^{\tau_2}{\left [j\f{(t-t')^2}{4}+f(t')\f{t-t'}{2}\right] 
dt'} + \omega(\tau_2-\tau_1)\leq 0.$$
or$$
j\f{(t-\tau_1)^3-(t-\tau_2)^3}{6}+2\omega(\tau_2-\tau_1)\leq -\int_{\tau_1}^{\tau_2}{f(t')(t-t')
dt'}.\eqno(26)$$
In particular (26) means that there are exist two close $\tau_1,\tau_2$ (such that $\delta=\tau_2
-\tau_1\ll t$), which satisfy this inequality. Then (26), by keeping only linear in $\delta$ 
terms, can be reduced approximately to$$
j\f{t-\tau}{2}+2\f{\omega}{t-\tau}\leq -f(\tau),\ {\rm where}\ \tau=(\tau_2+\tau_1)/2.\eqno(27)$$
Eq.(27) can be satisfied even on the initial time interval $t<T_1$. 

This situation is illustrated in Fig.7 where the current $j=1.5$ is formed by the flow with
the initial velocity only $\omega=0.3$.

\vskip0.0cm\hskip0.2cm
\epsfig{file=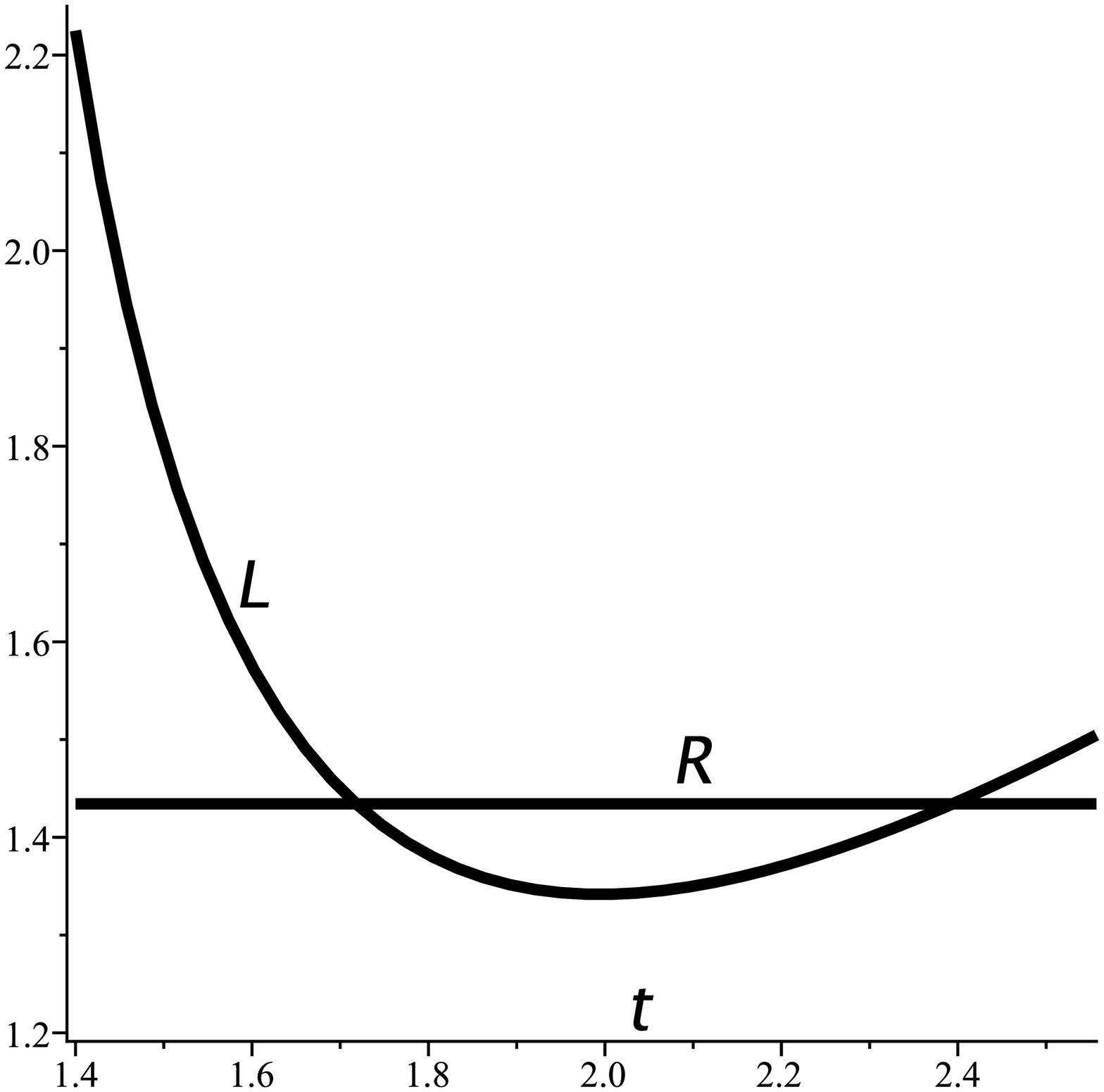, width=8cm, height=7cm} 

\vskip0.2cm
\centerline{\small FIG.7. Left L and right R sides of Eq.(27) for electrons emitted at $\tau=1.1$}
\medskip\noindent
In this case $T_1=2.557$ and when, for example $\tau=1.1$, Eq.(27) holds for the emission time on 
the interval $1.022<\tau<2.37$. If $\tau=1.022$ the curve only touches the line $-f(\tau)$ at 
$t=1.86$. In the case $\tau=1.1$ in Fig.7 the electrons pass some of them emitted earlier when 
$1.7<t<T_1$. Such regime cannot be treated by our method. The violation of our assumption on this 
subject takes place also for $\omega$ significantly larger than $0.3$, especially for $t>T_1$, we 
show here only a simplest situation. Note that the left side of Eq.(27) has its minimum 
$2\sqrt{j\omega}$ when $t-\tau=2\sqrt{\omega/j}$ which means that if $$
-f(t)<2\sqrt{j\omega},\ \ 0<t<\infty \eqno(28)$$
the technique of this work is valid.

Otherwise the inequality (27) can hold for some $t,\ \tau$ where the point $(x,t)$ 
in the Eulerian variables corresponds to at least two different points $(\tau_1,t)$ and 
$(\tau_2,t)$ when the flow is described by the Lagrangian coordinates. This undermines our method
for the case with $j=const$.

\bigskip
\noindent {\bf Discussion}
\bigskip				
 
In conclusion we note that the approximate treatment of the large scale flow variation developed
above is based on computations within short time intervals $T_1,T_2,...$. In each of 
them we have five equations which make possible the approximate evaluation of four flow 
parameters and provide the IC for the next interval. Computations on the first time interval 
$0<t<T_1$ can be done in closed form using Eq.(12) which makes them more precise especially in 
the case of an injected current $j=const$. Our techniques is illustrated only for the case when
the system was turned on, but its generalization for any given initial state is straightforward:
exactly as it was done in the paper in transitions between the time intervals $\Theta_{n-1}$ and
$\Theta_n$. The field emission model is more difficult: errors of computations are larger on the 
first interval where $f(t)$ varies significantly and Eq.(12) is not very helpful, the results for 
later times are more accurate. The limitations of our techniques are outlined and some estimates 
are done. The method of flow evolution is realized in our paper for an externally injected 
current and for a simple model of the field emission law, but it clearly can be extended to more 
realistic emission dependences and we are working presently in this direction.

\bigskip

\noindent {\bf Mathematical supplement}

\bigskip

Evolution of the cathode electric field in the case of externally injected current is described
by Eq.(23) which can be solved analytically for $t\leq T_1$. By differentiating (23) we have$$
\f{d^2f}{dt^2}(t)-\f{jt}{2}f(t)= \omega j+\f{j^2t^2}{4}\equiv h(t).\eqno(S1)$$
If $x_a=V_a=1$ the IC for $f(t)$ found earlier$$
f(0)=1,\ \ f'(0)=-j.\eqno(S2)$$	
Using standard procedures [11] we find [12] two independent solutions of the homogeneous equation 
$$
f_1(t)=\sqrt{t}K_{1/3}\left(\sqrt{2j}t^{3/2}/3\right),\ \ f_2(t)=\sqrt{t}I_{1/3}\left(\sqrt{2j}t^{3/2}/3\right),$$
where $K_\nu(z),I_\nu(z)$ are the modified Bessel functions, whose Wronskian is $$
W\{K_\nu(z),I_\nu(z)\}=1/z.$$
By a simple calculation we find$$
W\{f_1,f_2\}=f_1(t)f'_2(t)-f_2(t)f'_1(t)=3/2.\eqno(S3)$$
Using (S3) the solution of Eq.(S1) can be written [11] in the form$$
f(t)=C_1f_1(t)+C_2f_2(t)+\f{2}{3}\int_0^t{[f_2(t)f_1(s)-f_1(t)f_2(s)]h(s)ds}.\eqno(S4)$$
We satisfy the IC (S2) for $V_a=1$ by taking$$
C_1=\f{\sqrt{3}\Gamma(2/3)}{\pi}\left(\f{j}{18}\right)^{1/6},\ \ 
C_2=\Gamma(2/3)\left(\f{j}{18}\right)^{1/6}-\Gamma(4/3)(18j^5)^{1/6}.$$
Calculation show that $-f(t)$ can be large if $\omega$ is small. These Maple calculations by 
the polynomial approximation for Eq.(23) are faster than using the exact Eq.(S4).
Both functions $f_1$ and $f_2$ are non-negative and $t\geq s$ in (S4). It easy to show that 
the integral term in (S4) is positive: $f_2(t)/f_1(t)-f_2(s)/f_1(s)\geq 0$ because the ratio 
$f_2(t)/f_1(t)$ is an increasing function as its derivative has the same sign as $W\{f_1,f_2\}$,
see (S3). When $V_a=0$ we have$$
C_1=0,\ \ C_2=-\Gamma(4/3)(18j^5)^{1/6}.$$
This is the case studied in [13] by numerical simulations where however $j(t)$ increased linearly 
before becoming constant.

Note that Eqs.(24) for the case $j=const$ can be supplemented with an equation similar to (12)
but for an arbitrary discrete step $0<t<T_k$ (where $t$ is shifted on the interval $\Theta_{k-1}$): $$\f{df}{dt}=j(\omega T_k-1)+\f{jT_k}{12}(T_k^2-3tT_k+3t^2)+\f{j}{2}\left[\int_0^{T_k}{t'f(t')dt'}-
T_k\int_t^{T_k}{f(t')dt'}\right].\eqno(S5)$$
Obviously the transition time $T_k$ is a function of $t$. Eq.(S5) can be used for improving 
the accuracy of calculations, as well as a corresponding equation for the field emission case 
which it is more complicated. This was not done in our work because $f(t)$ variations for 
the later intervals are quite weak and the precision of modeling them by Eqs.(25) was already
better than for $t<T_1$.

\bigskip\noindent
{\bf ACKNOWLEDGMENTS}
\bigskip 

Work was supported in part by AFOSR Grant No. FA 9550-10-1-0131 and by NSF Grant DMR-1104501. 
The motivation of this study was initiated by our discussions with N.J. Fish to whom we 
express our thanks. We also benefited from discussions with J. Luginsland.

\end{document}